\documentclass[letterpaper]{article}
\usepackage[affil-it]{authblk}
\usepackage{setspace}
\usepackage{float}
\usepackage{graphicx}
\usepackage{physics}
\graphicspath{ {./figures/} }
\usepackage{subcaption}
\usepackage{xcolor}
\usepackage[normalem]{ulem}
\providecommand{\keywords}[1]

\usepackage[T1]{fontenc}
\usepackage{geometry}
\usepackage{setspace}
\usepackage{achemso}

\title{An Investigation into the Effect of Mobile Ions on the Steady State Performance of Perovskite Solar Cells }

\author[1,$\dagger$,*]{Matthew V. Cowley}
\author[2]{Kjeld O. Jensen}
\author[3]{Matthew J. Wolf}
\author[4]{Alison B. Walker}
\author[1]{Petra J. Cameron}

\affil[1]{Department of Chemistry, University of Bath, United Kingdom}
\affil[2]{Department of Mathematical Sciences, University of Bath, United Kingdom}
\affil[3]{Institute of Physical Chemistry, RWTH Aachen University, Germany}
\affil[4]{Department of Physics, University of Bath, United Kingdom}

\affil[*]{Corresponding author: m.cowley@ucl.ac.uk}
\affil[$\dagger$]{Present address: Institute of Infection, Immunity and Transplantation, University College London, United Kingdom}

\begin{document}
\date{}
\maketitle
\begin{abstract}
\noindent
 In perovskite solar cells, the interplay between mobile ions and photo-excited charge carriers is complex, with recent studies suggesting mobile ions can be either beneficial or detrimental to cell efficiency depending on the cell properties. In this study we use drift diffusion modelling and factorial analysis to simulate 601 pairs of perovskite solar cells (1202 total cells) across a broad range of physically relevant materials parameters. In each pair of devices, one cell contains mobile ions that can move freely. The second paired cell is identical but has no mobile ions. This approach allows us to systematically investigate the impact of mobile ions on cell performance. We deconvolve the contributions of key cell parameters including ion density, recombination rate and band offsets, for n-i-p and p-i-n devices with both organic and inorganic contact layers. Importantly, we find that in efficient devices, mobile ions have only a small impact on steady-state performance.

\end {abstract}


\section{Introduction}

High efficiency ($>27\%$) perovskite solar cells (PSCs) are made with defective solution processed lead halide perovskite thin films. These defects include ion vacancies, with halide ions having low activation energies for migration. This makes ions highly mobile within the perovskite film.\cite{Eames15, thiesbrummel_ion_2026} 
Mobile ion species in perovskite absorber materials are responsible for many phenomena observed in the opto-electronic measurements of perovskite solar cells (PSCs). For example they can cause current density-voltage (J-V) hysteresis: where the device history alters the J-V curve as the applied bias $V$ is varied from open circuit to short circuit and back to open circuit \cite{Garcia-Rodriguez22, Clarke23b, Jacobs17}. However, the extent to which mobile ions affect device characteristics, namely the power conversion efficiency, PCE, open circuit voltage ($V_{\mathrm{oc}}$), short circuit current ($J_{\mathrm{sc}}$) and fill factor (FF), and under what conditions, continues to be a matter of debate \cite{Cachafeiro25, thiesbrummel_ion_2026, Hart24}. A widely discussed effect of mobile ions is a reduction in current $J$, particularly close to short circuit. This happens when the mobile ions screen the net field across the perovskite layer that can otherwise aid electron and hole extraction \cite{Richardson16, Courtier19c,Clarke23a, Cave20}. The reduction in $J$ is enhanced when carrier mobilities are low \cite{Wu20} or carrier lifetimes are short, reducing electron and hole extraction into the charge transport layers. Recent papers have also linked an increase in mobile ion density that occurs during cell ageing to increased ion induced field screening and suggested that this is responsible for large efficiency losses in aged cells \cite{Thiesbrummel21, Thiesbrummel24}. On the other hand, there are a small number of studies which show that mobile ions can improve PSCs. Mobile ions in the perovskite layer can reduce the sensitivity of the photovoltage to energetic misalignments at the perovskite/transport layer interfaces. This happens because the redistribution of ions at the interface of the perovskite film with the charge transport layer can lower surface recombination currents at steady-state, increasing the photovoltage by 10-100 mV \cite{Hart24,Hill23,Cachafeiro25}.

Despite this, ion migration is frequently blamed for the majority of losses in PSCs. As recently highlighted by Cachafeiro and Tress \cite{Cachafeiro25} great care must be taken not to overestimate ionic losses. Experimentally, the influence of mobile ions on the device characteristics strongly depends on the device preconditioning protocol and the scan rate. In the case of a $J-V$ measurement this includes the value of $V$ and the illumination level applied when the cell is at steady-state before the start of the voltage sweep \cite{Hart24,Habisreutinger18,Cachafeiro25}. If any ionic redistribution (e.g. accumulation or depletion at the interfaces) occurs at the preconditioning voltage, it is incorrect to assume that a fast voltage scan will be an accurate measure of the ``ion-free" performance \cite{Cachafeiro25}. In fact, this is the case for all cells where the applied pre-conditioning potential is not exactly equal and opposite to the potential drop across the perovskite film \cite{Hart24}. Experimentally the applied voltage where the perovskite bulk is field free is typically measured to be a few hundred mV below the $V_{OC}$ \cite{Hill23, Hart24}, even for very high efficiency cells. In many studies an accurate value is not measured experimentally and is simply assumed to be close to the $V_{OC}$, meaning standard preconditioning conditions (typically close to or above $V_{OC}$) lead to ionic redistribution.

Due to the complex and varying influence of ionic redistribution during dynamic measurements, and because solar cells typically operate under steady state conditions, the steady state performance of PSCs is the most relevant performance metric for making design decisions. However, this can be difficult to measure for real devices as changes in ion distribution can occur over timescales of seconds, to hours or even months, making true `steady state' parameters hard to ascertain \cite{Habisreutinger18} experimentally. An additional problem is that some steady-state studies of PSC entirely neglect the contribution of mobile ions to the measured cell parameters, despite the fact that the steady state ionic distribution strongly impacts the measurement result \cite{Gong21}. 

Here a high-throughput computational approach has been used to investigate the effect of ion distribution on steady state cell parameters and hence to make design recommendations to minimize the impact of mobile ions on device performance. The advantage of computational device models is that they can be used to systematically investigate the parameter space. In order to make general observations of the influence of mobile ions on the performance of PSCs it is necessary to vary contact, transport layer, interfacial layer, and perovskite properties. Experimentally, this makes the number of material permutations very large. A Design of Experiment (DoE) approach is necessary as one-variable-at-a-time experimental designs only explore limited regions of parameter space, making it difficult to make general observations. Expanding to a multivariate experimental study is experimentally costly in terms of time and consumables. Computational studies using models at the appropriate level of theory for the problem make multi-variate exploration feasible \cite{LeCorre21a}. Additionally, studies using statistical designs to sparsely sample parameter space enable the study of very complex systems \cite{Baker21, Murray16}.

\section{Methods}

\subsection{Design of the Study}
In this study we use the drift diffusion software IonMonger \cite{Courtier19a, Clarke22} to simulate the impact of mobile ions on a broad range of experimentally relevant PSC configurations. This is combined with an efficient `design of experiment' (DoE) approach to study the effect of 32 different model parameters (a full list of the 32 parameters is given in Table \ref{tab:param_tab}) \cite{D0E00149J}. In the first step the DoE assigns values for the 32 variable model parameters. In the next step two simulations are run using those variables.  In the first simulation ions are allowed to move with the field. In the second simulation all the 32 variables are the same, but the mobile ions are evenly distributed across the perovskite film and fixed into position, they are not allowed to move. This equates to a material with a flat distribution of immobile vacancies distributed throughout the crystal lattice. We effectively compares two scenarios - a modelled device with mobile ions and a modelled device without mobile ions.  This method allows us to unambiguously ascertain the impact of mobile ions on the behaviour and efficiency of the modelled devices.

To ensure that the results are relevant to experimentally viable devices, all of the model parameters are guided by ranges of experimental values obtained from the literature (see Table \ref{tab:param_tab} for a complete list). This ensures that only realistic solutions to the drift diffusion model are obtained. The parameters are sourced from measurements on PSC with different absorber and contact layers. Here we look at built in voltages ($V_{bi}$) of between 1.35 and 1.15 eV and all of the simulated devices have Ohmic contacts.

A two-level parameter approach is taken, meaning, for each variable, a `high' and `low' value are chosen as the two possible settings, again as far as possible all of the high and low values chosen were based on experimentally measured values. This allows us to explore a large portion of parameter space efficiently. To exemplify, `high' and `low' values of the activation energy for ion migration are taken as 0.5\,eV and 0.3\,eV respectively; `high' and `low' values of the perovskite film thickness are set at 500\,nm and 300\,nm; permittivity values for the electron and hole transporting layers are set to `high' and `low' values of 10 $\epsilon_{0}$ and 3 $\epsilon_{0}$ which covers cells with organic electron and hole transporting layers, cells with inorganic hole and electron transporting layers and cells with a mixture of organic and inorganic layers. 

In this study the ion density is varied between 10$^{17} cm^{-3}$ and 10$^{19} cm^{-3}$ (10$^{23} m^{-3}$ to 10$^{25} m^{-3}$)\cite{mr3l-jg9h, Ehrlerhowmanyions}. The ion density in pristine cells may be lower than this, with values of 10$^{15} cm^{-3}$ to 10$^{17} cm^{-3}$ suggested in \cite{Thiesbrummel24}. However, the same study showed that the ion density increases rapidly as cells are operated under illumination, with MAPI cells seeing an increase to 10$^{16} cm^{-3}$ - 10$^{17} cm^{-3}$  after seconds to minutes of operation and more stable mixed cation cells seeing an increase to 10$^{17} cm^{-3}$ after 30 minutes to a few hours of operation \cite{Thiesbrummel24}. By selecting a minimum ion density of 10$^{17} cm^{-3}$ we ensure that the results are relevant to the longer term operating conditions of real cells and that the simulations compare the theoretical `ion free' PSC to PSC with reasonable to severe ion concentrations.

 To simulate an increase in defects which act as recombination centres, bulk SRH recombination was treated as a variable with the electron and hole lifetimes (SRH pseudo lifetimes) each varied by one order of magnitude. To simulate worsening interfaces, interfacial recombination velocity was varied over three orders of magnitude from 0.01 to 1 $ms^{-1}$. Recent modelling studies have looked at larger ranges of interfacial recombination, with top values of 1000 $ms^{-1}$ \cite{Thiesbrummel24}, but we chose to limit to the range expected for high efficiency devices.  

Once the parameters are selected, the simulated device is evaluated by running a `steady state' $JV$ measurement protocol. Experimentally this is equivalent to an ultra-slow  (10 $\mu Vs^{-1}$) experimental scan rate where the ions are allowed to reach their steady-state distribution at the applied voltage before the current is measured. In this situation, the maximum power point of the $JV$ curve corresponds exactly to that obtained in a steady state maximum power point measurement. It is important to emphasise that as a result, the shape of the $JV$ curves presented here are not dependent on ion dynamics, the ions are always at steady state for the voltage applied.

Due to the large number of permutations of 32 variables ($2^{32}$), a sparse sampling strategy is taken and correlations are used to predict ionic impact and find the parameters that are most significantly influenced by the ionic distribution under operating conditions. These results are used to understand the consequences of mobile ions and to gain insight into how an experimentalist can avoid conditions where the mobile ions are detrimental. Below, we explain our use of Design of Experiments and statistical tests, then present our results and their implications for the device physics before concluding with a discussion of how to avoid the detrimental impacts of ion migration and harness the benefits of ionic solar cells.

\subsection{Simulations}

\subsubsection{Theoretical treatments of ionic distributions}
To elaborate on our approach using `paired' simulated devices, we investigate the interaction between a property of a device and mobile ions by creating two copies of a device. One copy has \textbf{Dynamic Electric-field-driven Ions} - ions that are allowed to move with the field- and is denoted `DEI' in the following discussion. The second copy has \textbf{Site-bound Uniform Ions} -ions are uniformly distributed and fixed in place- and is denoted `SUI'. This corresponds to a theoretical mobile ion free device. As stated above, ions in the DEI case are allowed to freely move according to the drift-diffusion model \cite{Nelson03}. This represents the likely behaviour in most perovskite devices, where due to the influence of the electric field and the bias history, the concentration of iodide vacancies $N_{V_I}$ is a function of position $x$, applied voltage $V_{\mathrm{app}}$, and time $t$,\cite{Jacobs17}
\begin{equation}
    N_{V_I,t} = f(x, V_{app}, t).
\end{equation}
We eliminate the time component of  $N_{V_I,t}$ by employing a scan rate that is much slower than the timescale for ion motion across the Debye layer, the region in the perovskite layer adjacent to the hole transport layer where the ions accumulate. This allows complete relaxation at every applied voltage. As long as the scan rate is slow enough, the shape of the $JV$ curve becomes scan rate independent. Therefore,
\begin{equation}
    N_{V_I} = f(x, V_{app}).
\end{equation}

In contrast in the the SUI case, the ions are immobile and fixed in a uniform distribution, representing a device where ions cannot move or accumulate. This is a hypothetical device, identical to the DEI in all aspects, except the activation energy for ion migration is so high that the ions are effectively fixed in their lattice sites such that $N_{V_I}$ is equal to the ionic density $N_0$ across the perovskite layer.

\subsubsection{Design of Experiments} 

\begin{figure}
    \includegraphics[width=\linewidth]{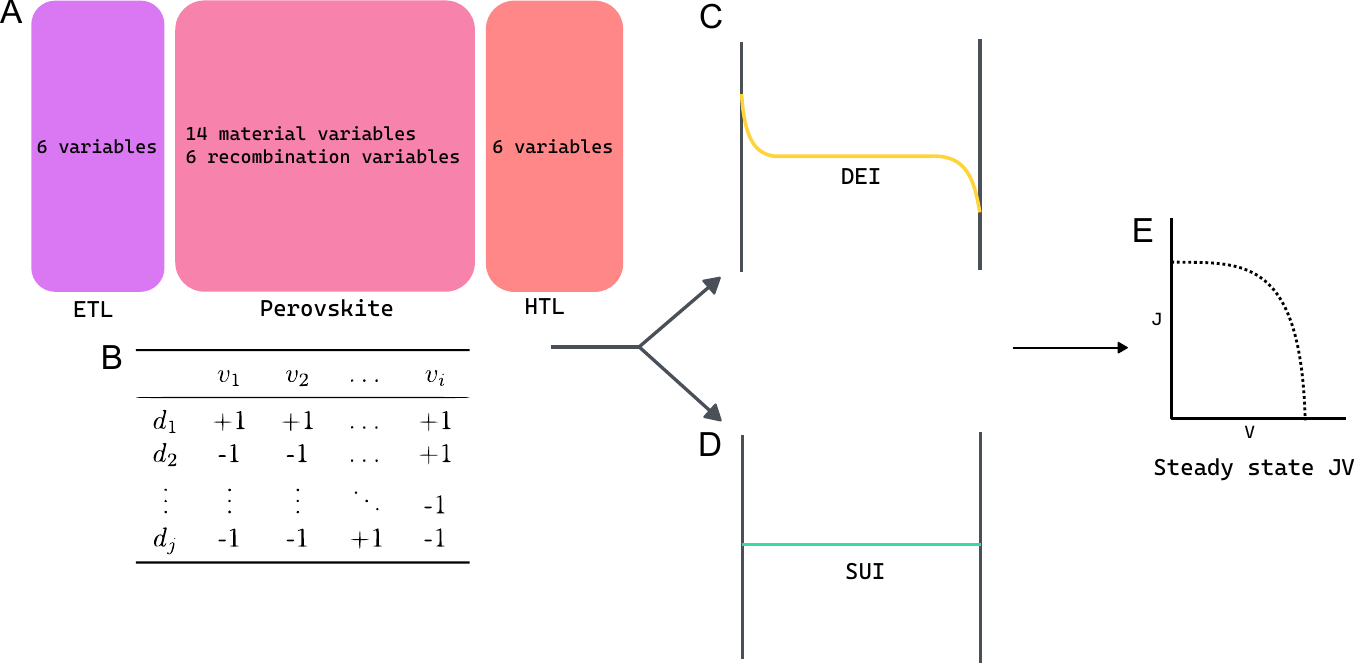}
    \caption{ 
    \textbf{An efficient simulation study to deconvolve the influence of parameters and conditions.}
    (A) Schematic showing the distribution of investigated parameters across the layers that comprise a perovskite solar cell, including the Electron Transport Layer, ETL and Hole Transport Layer, HTL.
    (B) Table illustrating the two-level factorial design of the study.
    (C) Schematic plots showing the two paradigms investigated: mobile ions are distributed according to dynamic equilibrium (DEI) (yellow) or (D) mobile ions are static in a uniform distribution (SUI) (Teal)  (E) Schematic diagram of the ultra-slow J-V scan performed to obtain the statistics reported .
    }
    \label{fig:outline}
\end{figure}

The approach of sparsely sampling a high dimensional parameter space is called Design of Experiments. Using the open source simulation software IonMonger (IM) \cite{Courtier18, Clarke22} as our probe, we solve the drift-diffusion equations for a three-layer PSC in one dimension (Figure~\ref{fig:outline}A). As described above, the parameter space of a three-layer PSC is large, making a two-level factorial design necessary to explore how parameters influence the difference between the two configurations. For example, a high (+1) and low(-1) value of each variable parameter is defined.(Figure~\ref{fig:outline}B).

As outlined above, the choices for high and low values are informed by the ranges of values reported experimentally for PSCs in the literature and are described in Table~\ref{tab:param_tab}. A decision was taken not to look at extremely broad parameter ranges as this often returns experimentally unrealistic devices. 

\begin{table}
    \footnotesize 
    \centering
    \caption{Table showing parameters used in the simulations. Parameters with two values are those included in the multivariate study.}
    \label{tab:param_tab}
      \begin{tabular}{llrrr}
      \hline
      Variable          & Description & Low ($-1$) & High ($+1$) & Justification     \\
      \hline & Perovskite \\
      \hline  $b^{pk}$ & Layer width (m) & 3$\times$10$^{-7}$ & 
      5$\times$10$^{-7}$ & \cite{Bag15,Tang18}    \\
      $\epsilon^{pk}$  & Permittivity (Fm$^{-1}$) & 15$\epsilon_0$ & 30$\epsilon_0$ & \cite{Brivio14,Futscher19}   \\
      $\alpha$     & Absorption coefficient (m$^{-1}$) & 1$\times$10$^{7}$  & 2$\times$10$^{7}$ & \cite{Loper15}  \\
      $E_{c}^{pk}$      & Conduction band minimum (eV)  & -3.8 & -3.75 & \cite{Schulz14} \\
      $E_v^{pk}$      & Valence band maximum (eV)   & -5.4 & -5.35 & \cite{Schulz14} \\
      $g_c^{pk}$      & Conduction band density of states (m$^{-3}$)  & 1$\times$10$^{24}$ & 1$\times$10$^{25}$  & \cite{Brivio14} \\
      $g_v^{pk}$      & Valence band density of states (m$^{-3}$)    & 1$\times$10$^{24}$ & 1$\times$10$^{25}$ & \cite{Brivio14} \\
      $D_n^{pk}$      & Electron diffusion coefficient (m$^{2}$s$^{-1}$)
       &  1$\times$10$^{-5}$ & 1$\times$10$^{-4}$ & \cite{Stranks13,stavrakasVisualizingBuriedLocal2020,agustinChargeCarrier2021,Hill23} \\
      $D_p^{pk}$      & Hole diffusion coefficient (m$^{2}$s$^{-1}$) 
      &1$\times$10$^{-5}$ & 1$\times$10$^{-4}$ & \cite{Stranks13,stavrakasVisualizingBuriedLocal2020,agustinChargeCarrier2021,Hill23} \\
      $N_I$           & Mobile ion density (m$^{-3}$) & 
      1$\times$10$^{23}$ & 1$\times$10$^{25}$ & \cite{Walsh15,Bertoluzzi20a} \\
      $I_{\mathrm{lim}}$     & Maximum local vacancy density (m$^{-3}$) & 
      1$\times$10$^{26}$ & 1$\times$10$^{28}$ & \cite{Walsh15,kressPersistentIonAccumulation2022} \\
      $\mathcal{D}_{\mathrm{I}}$  & High temp. ionic diffusion coefficient (m$^{2}$s$^{-1}$) & 1$\times$10$^{-7}$  & 1$\times$10$^{-6}$  & \cite{Richardson16,Cave20} \\
      $E_A$  & Activation energy of ion migration (eV) & 0.3 & 0.5 & \cite{Eames15} \\
      $l$  & Normal or inverted architecture (boolean) & \texttt{False} & \texttt{True} & \cite{Garcia-Rodriguez22} \\
      $\tau_n$        & SRH electron pseudo lifetime (s) & 1$\times$10$^{-8}$  & 
      1$\times$10$^{-7}$ & \cite{DeQuilettes15,Sherkar17,Kirchartz19,Duijnstee20,Ni20,yuanDerivingMobilitylifetimeProducts2025} \\
      $\tau_\mathrm{p}$        & SRH hole pseudo lifetime (s) & 1$\times$10$^{-8}$ & 1$\times$10$^{-7}$  & 
      \cite{DeQuilettes15,Sherkar17,Kirchartz19,Duijnstee20,Ni20,yuanDerivingMobilitylifetimeProducts2025} \\
      $\nu_n^{el}$    & SRH electron surface velocity at ETL (ms$^{-1}$)  & 0.01 & 1 & \cite{Kirchartz19,Jariwala21,Yang17} \\
      $\nu_p^{el}$    & SRH hole surface velocity at ETL (ms$^{-1}$)   & 0.01 & 1 & \cite{Kirchartz19,Jariwala21,Yang17} \\
      $\nu_n^{hl}$    & SRH electron surface velocity at HTL (ms$^{-1}$)   & 0.01 & 1 & \cite{Kirchartz19,Jariwala21,Yang17} \\
      $\nu_p^{hl}$    & SRH hole surface velocity at HTL (ms$^{-1}$)   & 0.01 & 1 & \cite{Kirchartz19,Jariwala21,Yang17} \\
      $\beta$   & Bimolecular coefficient (m$^3$s$^{-1}$) & 2$\times$10$^{-17}$
      & N/A & \cite{DeQuilettes16,Kirchartz19} \\
      $C_{n,p}$         & Auger coefficient (m$^6$s$^{-1}$) & 4.0$\times$10$^{-40}$
      & N/A & \cite{Crothers17,DeQuilettes16} \\
      \hline & Electron transport layer \\
      \hline
      $b^{el}$ & Layer width (m) & 1$\times$10$^{-8}$ & 1$\times$10$^{-7}$& 
      \cite{Salim18,Tang18,Li18c,Mukametkali23,Luo19} \\
      $N_D^{el}$      & Donor doping density (m$^{-3}$)  & 5$\times$10$^{21}$ & 5$\times$10$^{22}$ &
      \cite{Sellers11,Balderrama15,Garcia-Belmonte08,Liu15,Shang16,Hochgesang22,Abate14} \\
      $g_c^{el}$  & Conduction band density of states (m$^{-3}$) & 1$\times$10$^{24}$ & 1$\times$10$^{25}$ &
      \cite{Zhang14,Garcia-Belmonte10,Spear1973} \\
      $E_\mathrm{c}^{\mathrm{el}}$      & Conduction band minimum (eV) & -4.0 & -3.9 & 
      \cite{Schulz14,Isikgor22,Mikroyannidis11,He10}\\
      $D_n^{el}$      & Electron diffusion coefficient (m$^2$\,s$^{-1}$) & 1$\times$10$^{-9}$ & 1$\times$10$^{-5}$ & 
      \cite{Mi15,Nakka22,Armin17,Wang12,Krasienapibal14} \\
      $\epsilon^{el}$ & Permittivity (Fm$^{-1}$) & 3$\epsilon_0$ & 10$\epsilon_0$ & 
      \cite{Wypych14,Balderrama15,Rao1965,Stamate03} \\
      \hline
       & Hole transport layer \\
      \hline
      $b^{hl}$ & Layer width (m)   & 1$\times$10$^{-8}$ & 1$\times$10$^{-7}$ & \cite{Salim18,Tang18,Li18c,Mukametkali23,Luo19} \\
      $N_A^{hl}$      & Acceptor doping density (m$^{-3}$) & 
      5$\times$10$^{21}$ & 5$\times$10$^{22}$ & 
      \cite{Sellers11,Balderrama15,Garcia-Belmonte08,Liu15,Shang16,Hochgesang22,Abate14} \\
      $g_v^{hl}$      & Valance band density of states (m$^{-3}$) & 1$\times$10$^{24}$ & 1$\times$10$^{25}$ & 
      \cite{Zhang14,Garcia-Belmonte10,Spear1973} \\
      $E_\mathrm{v}^{\mathrm{hl}}$      & Valence band maximum (eV)  & -5.25 & -5.15 & 
      \cite{Schulz14,Isikgor22,Koffyberg1981,Nakaoka04,Ratcliff11} \\
      $D_p^{hl}$      & Hole diffusion coefficient (m$^2$\,s$^{-1}$) & 1$\times$10$^{-9}$ & 
      1$\times$10$^{-5}$ & \cite{Mi15,Nakka22,Armin17,Wang12,Krasienapibal14} \\
      $\epsilon^{hl}$ & Permittivity (Fm$^{-1}$) & 3$\epsilon_0$ & 10$\epsilon_0$ &      \cite{Wypych14,Balderrama15,Rao1965,Stamate03}  \\
      \hline
    \end{tabular}
\end{table}

A full two-level factorial design would require $2^N$ experiments, which for the 32-factor case (2$^{32}$ = 4.295 $\times$ 10$^{12}$) would be computationally prohibitive; therefore, a fractional design was used. Here, 2$^{10}$, i.e. 1024, `devices' are sampled from the parameter space to study the main effects that dominate the system, reducing the number of simulations by a factor of 2$^{22}$, i.e. 4.19$\times$10$^6$.
In the terminology of experimental design, this corresponds to a resolution of 9, a design which allows the effect of a single factor to be analysed separately from a combination of up to 8 factors \cite{Lazic04}.

The resulting 1024 modelled DEI and 1024 SUI `devices' ($d_1$, $d_2$, \dots, $d_j$) are selected in such a way that they have equal proportions of the high and low settings for each factor. It can be represented as selecting 1024 vertices of a 32-D hypercube. We ensure this sampling is extensive enough by checking the convergence in the correlation coefficients between each parameter and the ratio PCEs of the DEI and SUI case as the number of samples is increased (Figure~\ref{fig:convergence}).

To exemplify, as the transport layers in a perovskite can be either organic or inorganic materials---both classes with very different magnitudes for properties such as permittivity---the low value represents one class and the high value represents the other (e.g. a relative permittivity of 3 for organic semiconductors and of 10 for inorganic semiconductors). This aids in covering the vast space of possible design decisions, and helps the study provide a comprehensive picture of ionic impact in PSC.

For each parameter combination, a DEI and an SUI simulation was run with an ultra-slow scan rate, selected here as 1$\times$10$^{-5}$ Vs$^{-1}$ (Figure~\ref{fig:outline}C). We confirmed that this scan rate can be shown to be effectively steady-state by allowing the devices to come to equilibrium at the maximum power point (MPP) obtained from the slow JV scans for 1$\times$10$^5$ s. The final maximum power output $P_{max}$ of these simulations was found to deviate from $P_{max}$ of the slow JV scans by only 2.0$\times$10$^{-6}$\% on average, with no device showing a deviation greater than 1$\times$10$^{-4}$\% (Figure~\ref{fig:p_max_error}) giving confidence that the measured cells are indeed at steady state.

\subsubsection{Absorption models}
The simulations evaluate both $n$-$i$-$p$ and $p$-$i$-$n$ configurations as a parameter of the experimental design, with light entering through either the ETL or HTL. IonMonger assumes a Beer-Lambert absorption model, with a single wavelength-independent illumination intensity, $F_{\mathrm{ph}}$,  representing all photons that can potentially be absorbed in the perovskite layer. This means we must calculate the $F_{\mathrm{ph}}$ per device based on the perovskite bandgap (Figure~\ref{fig:eg_fph}), $E_g$, via, 
\begin{equation}
    F_{\mathrm{ph}}(E_g) = 0.9 L_S \int_{E_g}^{\infty} F_{AM1.5}(E) \rm{d} E.
\end{equation}
Here, $L_S$ is the light intensity in suns  and $F_{AM1.5}$ the standard AM1.5 solar spectrum photon flux for energies with energy equal to or above the device band gap, the factor of 0.9 accounts for a reflection loss of 10\% from the top surface(s) of the device \cite{Wang22}.

\subsubsection{Numerical limitations}
Although the full design consists of 1024 pairs of devices, the results presented here only represent 601 pairs of devices, e.g. 601 with mobile ions and 601 without ions. In order to compare the DEI and SUI conditions, both sets of parameters have to be solvable in IonMonger. A fraction of input combinations could not readily be solved, nearly all the failures being for SUI (Figure~\ref{fig:solved_bar_plot}).
This most likely is due to IonMonger's numerical design, where a number of choices have been made to ensure a greater degree of success with the existence of mobile ionic accumulation and large potential drops at interfaces \cite{Courtier18}.

Rewriting the model to allow for greater parameter diversity in the SUI case is beyond the scope of this work, and the remaining 601 pairs of runs are sufficient to describe the full range of device behaviour. This is evidenced by the fact that the correlation coefficients between each parameter of the model and the ratio of the PCEs of the DEI and SUI case have converged after sampling $\approx$\,600 devices (Figure~\ref{fig:convergence}). Additionally, the devices with unsolved SUI simulations have a very similar distribution of DEI PCEs when compared to the 601 DEI devices which were successfully modelled in both the SUI and DEI cases, with a slightly higher mean and a smaller variance (Figure~\ref{fig:kde_failed}). As the correlation between DEI and SUI PCE is very strong for all but the lowest PCE devices in the dataset, we can therefore assume that no significant variation in ionic behaviour within the parameter space is being omitted by excluding the devices where the SUI case failed to solve.

\section{Results and Discussion}

\subsection{Impact of Mobile Ions on Steady State Power Conversion} \label{sec:data_prop}
\begin{figure}
    \centering
    \includegraphics[width=\linewidth]{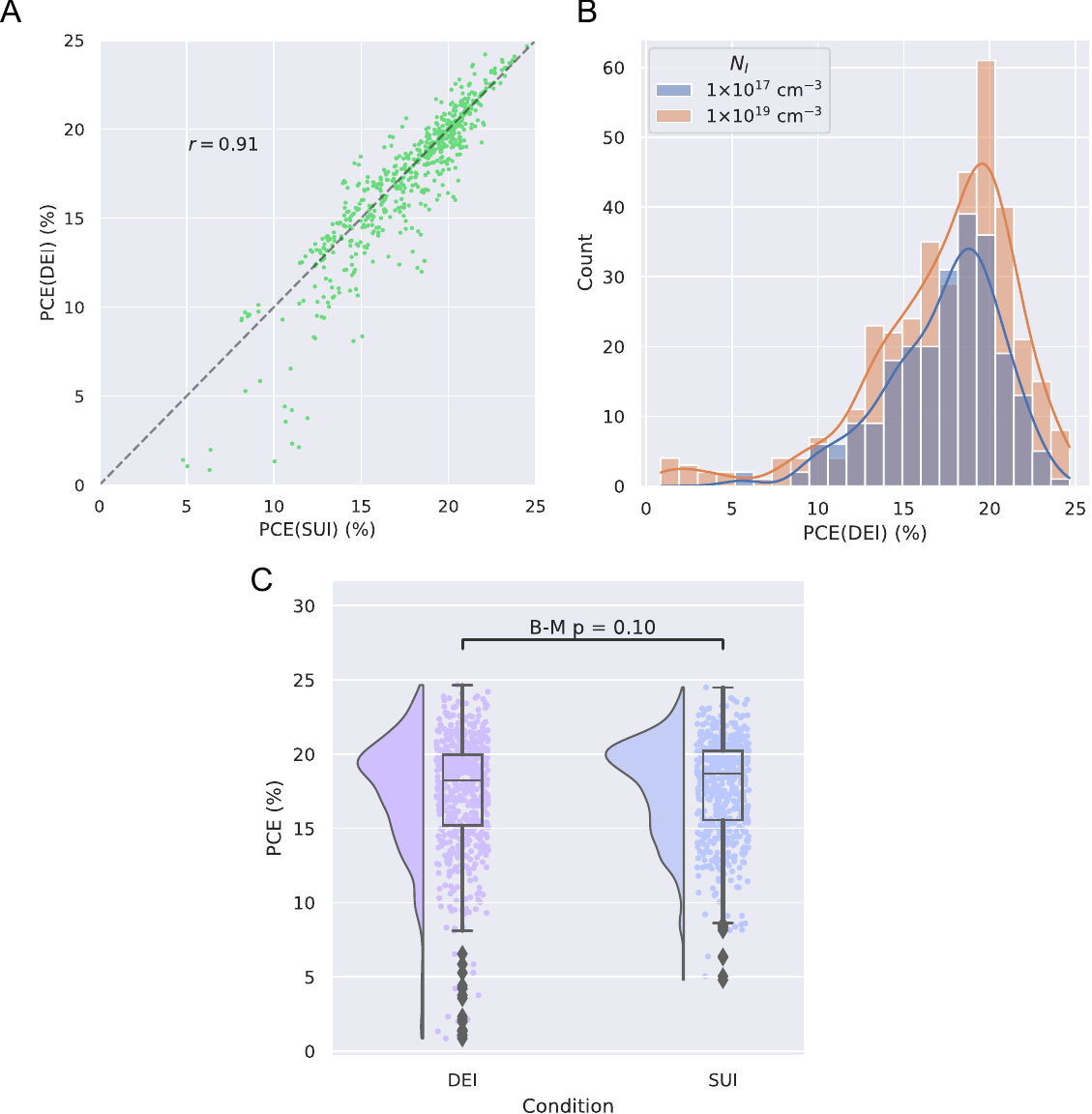}
   \caption{
   \textbf{The effect of mobile ions is minimal compared to the physically relevant parameter range.}
    (A)
    Scatter plot of power-conversion efficiencies (PCE) of parameter pairs of devices with mobile ions (DEI) and those without (SUI).
    The dashed line is a $y=x$ line for visual aid, and $r$ is Pearson's correlation coefficient.
    (B)
    Histogram and kernel density estimation of the PCE of DEI devices, stratified by equilibrium ion density $N_I$.
    (C)
    Raincloud plot showing density and swarm visualisations of the power-conversion efficiency (PCE) distribution of devices simulated either with mobile ions (DEI) or with static ions (SUI).
    Brunner-Munzel (B-M) test (unpaired) applied between the distributions.
    }
    \label{fig:data_distributions}
\end{figure}

To address our main question of whether mobile ions in perovskite solar cells affect operational steady state performance, we first plot the PCE of the SUI vs the PCE of the DEI case for a given parameter set in a scatter plot. E.g. for each pair of cells (601 pairs, 1202 total cells) the efficiency of the cell where ions are mobile is plotted versus the efficiency of the paired cell where all parameters are identical except that the ions are uniform, fixed and immobile. Interestingly we can see that the two measures are highly correlated (Figure~\ref{fig:data_distributions} A); i.e. that the efficiencies of paired SUI and DEI cells are generally similar. 

Two notable features are displayed.
First, poorly performing devices ($<$15\,\% PCE for the SUI case) are generally made worse by the inclusion of mobile ions, in rough proportion to how poorly they perform.
Second, excellent devices ($>$15\,\% SUI PCE) are minimally affected by the addition of mobile ions, and in some cases are improved. This is true both for the `low' ion concentration of 10$^{17}$\,cm$^{-3}$ and for the `high' ion concentration of 10$^{19}$cm$^{-3}$ (Figure~\ref{fig:data_distributions}B). It is worth emphasising that at the higher ion concentration in particular, the applied field is completely screened at each point of the steady state $JV$ curve. Despite this the ions do not reduce the steady state device efficiency of good cells, showing that an increase in ion concentration is not automatically linked with a degradation of cell parameters.

These observations are reinforced by comparing the results from the two ionic conditions (DEI and SUI) plotted as independent distributions (Figure~\ref{fig:data_distributions} C). 
The distribution of efficiencies for the simulations with mobile ions (DEI) is only slightly broader than the distribution of efficiencies of identical devices without ions. Additionally the DEI case has only a very slightly lower median. A statistical test can be used to quantify the difference between the simulated efficiencies from the DEI and the SUI cells. Due to the distribution of PCEs being non-normal, and there being unequal variance between the DEI and SUI conditions, the Brunner-Munzel test is used. This is a non-parametric (i.e. no underlying distribution is assumed) and heteroscedastic (i.e. no assumption is made of the two distributions having equal variances) test for independent samples \cite{Brunner00}. Importantly, this test indicates \textbf{no significant difference between the two distributions}. While this is not a demonstration of equivalence, this failure to reject the null shows that within the range of material property variation explored here, the impact of ions on steady-state PCE is not large enough to produce a statistically meaningful effect, despite an extensive range of device performance and behaviour being explored. To reiterate, this means that when the efficiencies of a population of 601 cells with mobile ions is compared to a population of 601 cells without mobile ions, it is not possible to differentiate the cells on the basis of whether ions are present. This observation suggests that experimentally, ion migration may often be prematurely blamed for a reduction in cell efficiency.

\subsection{Mobile ions can improve device performance}

Having examined the behaviour of the whole population of devices, we will now introduce ``ion-normalised" statistics as a method to understand the impact of mobile ions on the J-V characteristics of pairs of cells which are identical in all aspects except that one cell has mobile ions and one does not. 

This involves comparing PCE, J$_{sc}$,  $V_{oc}$ and FF for the DEI and SUI simulations i.e.,
\begin{equation}
    \tilde{S} = S_{\mathrm{DEI}} / S_{\mathrm{SUI}},
\end{equation}
where $S$ is a given J-V characteristic such as PCE or J$_{sc}$. For example, the ion-normalised PCE $\tilde{\eta}$ is
\begin{equation}
    \tilde{\eta} = \frac{\eta_{DEI}}{\eta_{SUI}}.
\end{equation}
As shown, for a single pair of SUI and DEI cells, this is simply the ratio of the cell efficiencies with ions and without ions. As such $\tilde{\eta}$ encodes information about the impact of mobile ions on cell efficiency. A calculated value of $\tilde{\eta} > 1$ represents a pair of devices where the PCE is larger with ions, $\tilde{\eta}\approx 1$ describes a pair of devices which have the same PCE, and $\tilde{\eta} < 1$ is a pair of devices where a cell without mobile ions is more efficient.

\begin{figure}
    \includegraphics[width=\linewidth]{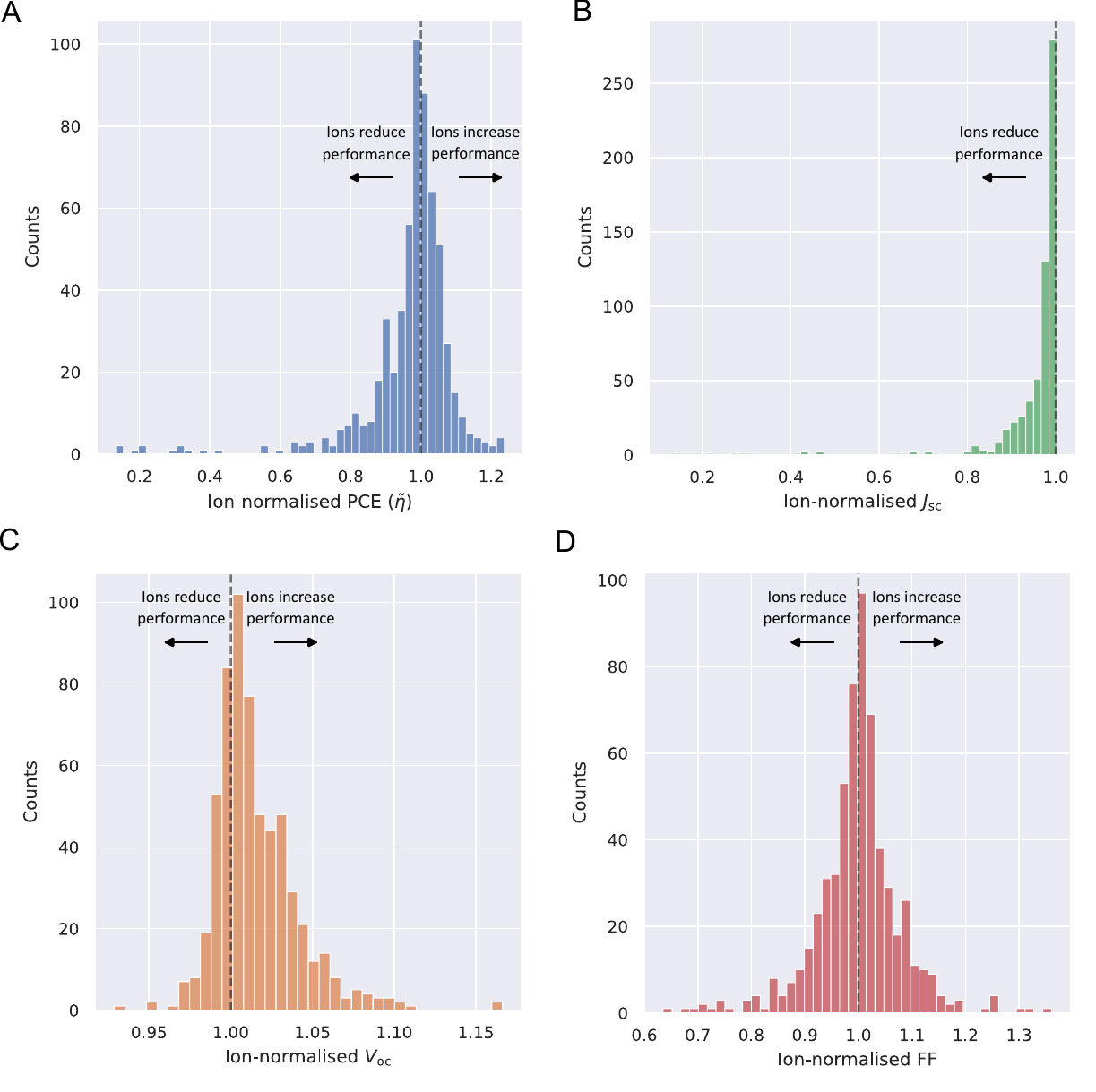}
    \caption{
    \textbf{Key statistics of solar cell quality are affected by ions.}
    Histograms of ion-normalised (mobile ion simulation / static ion simulation) (A) power-conversion efficiencies (PCE), (B) short-circuit current ($J_{{sc}}$), (C) open-circuit voltage ($V_{\mathrm{oc}}$), and (D) fill-factor (FF).
    }
    \label{fig:jv_statistics}
\end{figure}

Collectively, the four panels in Figure \ref{fig:jv_statistics} show that in most cases the presence of ions only causes a +/- 10\,\% change in the JV characteristics.
The distribution of $\tilde{\eta}$ in Figure \ref{fig:jv_statistics}A shows there is a left-skewed normal distribution centred around 1 (indicated by the dashed line on the graph). About one third of the cells show little change in efficiency in the presence of ions, one third show an improvement and one third show a reduction in efficiency in the presence of ions. Only a small number of poorly performing cells are made substantially worse by the presence of mobile ions. That only a minority of devices are severely impacted by the freely moving ionic distribution is key, as it means that for many parameter sets, the presence of mobile ions is only a minor factor in their operational performance, at least for cells with ion densities between 10$^{17}$\,cm$^{-3}$ and 10$^{19}\,$cm$^{-3}$. 

The right-skewed distribution for the ion-normalised steady-state $V_{\mathrm{oc}}$ (Figure \ref{fig:jv_statistics}C), shows that ions tend to increase $V_{\mathrm{oc}}$. It is well established in the literature that after preconditioning at a forward bias, the resulting ion distribution enhances the electric field and boosts $V_{\mathrm{oc}}$ \cite{Jacobs17, Shen17}, but here it is shown that this effect can persist under steady state operational conditions.

In contrast steady-state $J_{sc}$ is slightly degraded by the presence of mobile ions in all device cases studied here (Figure \ref{fig:jv_statistics}B).
This happens only by a small amount (less than 10\,\%) for the majority of devices and the high count density near 1 shows that in a large number of cases, there is little change between the DEI and SUI conditions.
The reported distribution agrees with results in the literature showing universal short-circuit current losses in PCE resulting from ion redistribution \cite{Hart24}.

The distribution of FF is almost perfectly normal, with the histogram bin just greater than 1 having the highest counts (Figure \ref{fig:jv_statistics}D).
In the literature, the presence of mobile ions is generally considered to lower the FF. This is attributed to similar effects as the losses in $J_{sc}$, where the ions screen the electric field and reduce the drift current \cite{Wu20}. There are also reports of the FF being improved, but only in instances when the applied voltage on the cell is such that the halide vacancies are no longer accumulated at the HTL, but are instead at the ETL. This situation is discussed from the perspective of forward pre-biasing by Jacobs et al. \cite{Jacobs17} Here we show that both improved and degraded fill factors are observed in the steady-state $JV$ curves. 

\subsection{The connection between transport layer band positions, hole lifetimes and device resilience} 
\label{sec:vbi_recomb}

\begin{figure}[tb]
    \includegraphics[width=\linewidth]{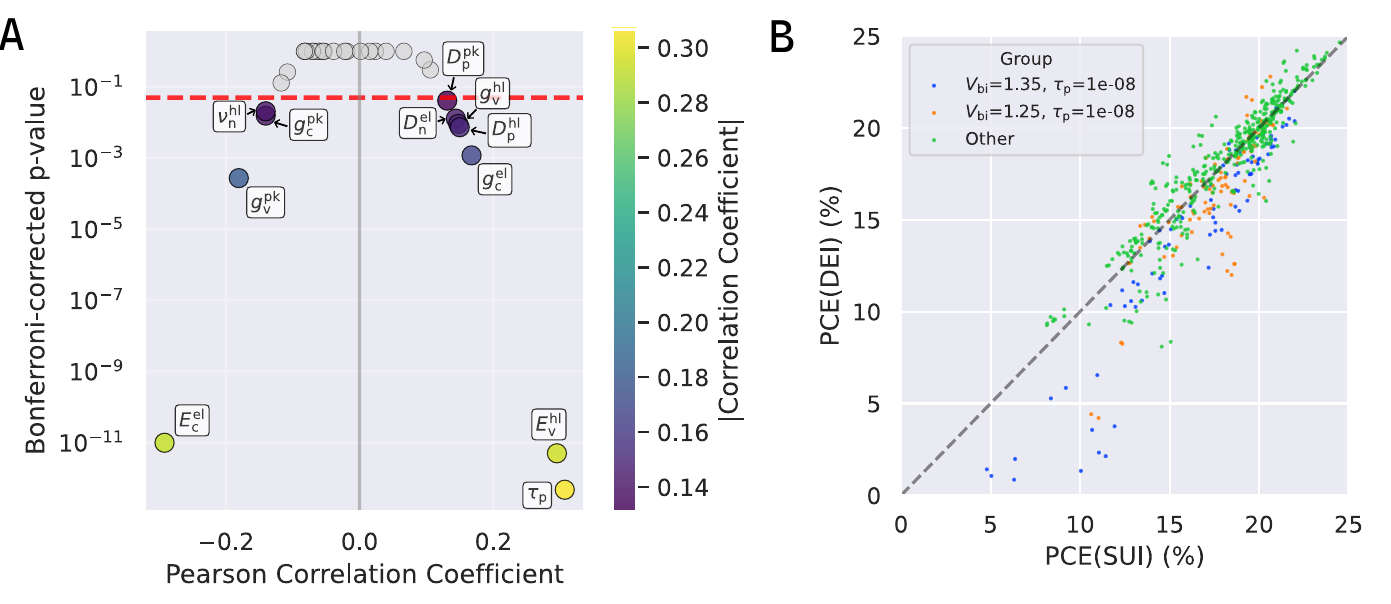}
    \caption{
    \textbf{Engineering a small built-in voltage and a long hole lifetime is protective against the presence of mobile ions.}
    (A) Scatter plot of the results from calculating Pearson's correlation coefficient between each of the 32 model variables and the $\tilde{\eta}$.
    The Bonferroni corrected p-value ($p*N$ where $N$ is the number of comparisons (32)) is plotted against Pearson's $r$.
    Grey dots represent non-significant correlations.
    Coloured dots represent significant correlations, with intensity determined by the absolute magnitude of the correlation $|r|$.
    (B) Scatter plot of PCE of parameter pairs of DEI and SUI devices.
    Dots are coloured by their values for $V_{\mathrm{bi}}= E_\mathrm{c}^{\mathrm{el}} - E_\mathrm{v}^{\mathrm{hl}}$ and $\tau_\mathrm{p}$.
    }
    \label{fig:correlations}
\end{figure}

The above data shows that the majority of devices are unaffected by the presence of mobile ions at steady state.
However, there is a subset of devices that are severely negatively affected by transitioning from the SUI case to the DEI case.
To investigate the underlying mechanisms systematically, we computed the Pearson correlation coefficient of each variable with $\tilde{\eta}$ (Figure~\ref{fig:correlations}A).
The results are plotted in variant of a `volcano plot'; these plots are commonly used in biology, but may be unfamiliar to a chemistry abd physics audience. In this plot, all of the model parameters are plotted according to how strongly they correlate with the value of $\tilde{\eta}$. The x-axis shows the Pearson correlation coefficient $r$ between each variable and $\tilde{\eta}$, while the y-axis shows the Bonferroni-corrected $p$-value. A smaller p-value indicates that the observed correlation is less likely to have arisen by chance (under the null hypothesis of no association). The Bonferroni correction limits the proportion of false discoveries by this approach to less than the significance threshold, $\alpha$, selected. It does this by making the threshold for significance more stringent by a factor equal to the number of tests. A larger $|r|$ indicates a stronger linear association between that variable and $\tilde{\eta}$. Variables toward the lower corners of the plot are the strongest candidates for a genuine relationship with mobile ion impact. To note, correlation strength should not be read as a direct measure of physical or causal impact, and p-value should not be interpreted as a probability of certainty.
Applying this analysis, 11 out of 32 features have a significant correlation ($p <\alpha = 0.05$) with $\tilde{\eta}$, with the highest correlations for the hole lifetime $\tau_{\mathrm{p}}$, as well as the energies of the valence band edge in the HTL and the conduction band edge in the ETL $E_\mathrm{v}^{\mathrm{hl}}$, and $E_\mathrm{c}^{\mathrm{el}}$.
These have an absolute magnitude of $r > 0.2$.

The strength of the correlation between $\tilde{\eta}$ and the transport layer band positions ($E_\mathrm{v}^{\mathrm{hl}}$, $E_\mathrm{c}^{\mathrm{el}}$) is likely tied to these parameters' influence on the built-in voltage of the device.
In the case of highly doped contact layers, the built-in voltage can be related to \cite{Courtier18},
\begin{equation}
    V_{\mathrm{bi}}= E_\mathrm{c}^{\mathrm{el}} - E_\mathrm{v}^{\mathrm{hl}}.
\end{equation}
The sign of $r$ is positive for $E_\mathrm{v}^{\mathrm{hl}}$ and negative for $E_\mathrm{c}^{\mathrm{el}}$, meaning that increasing $E_\mathrm{v}^{\mathrm{hl}}$ tends to increase $\tilde{\eta}$ and increasing $E_\mathrm{c}^{\mathrm{el}}$ tends to decrease $\tilde{\eta}$.
Therefore a large $V_{\mathrm{bi}}$ generally results in a more negative impact of mobile ions on performance. In common with \cite{Cachafeiro25} we find that if a device is relying on the internal field generated by a large $V_{\mathrm{bi}}$ for charge extraction, when mobile ions screen this field, the performance is significantly impacted.

The high importance of $\tau_\mathrm{p}$ is interesting as the impact of the spatial distribution of ions on the electric field and the presence of defect-mediated recombination mechanisms has previously been argued to be responsible for observed ionic effects \cite{Garcia-Rodriguez22}.
The sign of $r$ being positive shows that increasing the hole lifetime reduces the impact of the ions screening the internal field.
This is intuitive as the charge carriers are less sensitive to the reduced magnitude of drift as they have more time to diffuse out of the perovskite layer.
Together, these parameters suggest a model where, in the condition of poor hole lifetime in the perovskite, the band positions of the transport layers are crucial in mediating the effect of mobile ions.

That $V_{\mathrm{bi}}$ and $\tau_\mathrm{p}$ are key parameters can be clearly seen via a scatter plot of SUI and DEI PCEs (Figure~\ref{fig:correlations}B), where each point represents a pair of simulations with given device parameters.
The dots are coloured with respect to their parameters.
The blue dots represent devices that correspond to the worst-case scenario, having a large $V_{\mathrm{bi}}$ (as $E_\mathrm{v}^{\mathrm{hl}}$ is low, and $E_\mathrm{c}^{\mathrm{el}}$ is high) and a short hole lifetime in the perovskite $\tau_\mathrm{p}$.
Most of the poorly performing devices, both in terms of the absolute values of PCE, and the greatest impact of the mobile ions (PCE(DEI) well below PCE(SUI)), fall into this category.
Very few devices with these parameters see improvements or maintenance of performance from introducing mobile ions.
The orange dots correspond to devices that have a slightly smaller $V_{\mathrm{bi}}$ than the blue due to an increased value of $E_\mathrm{v}^{\mathrm{hl}}$.
These devices are slightly better off, but can still see large negative impacts from introducing mobile ions.
The majority of other parameters (green dots) fare much better.
As evidenced by the variance of PCE(SUI) and PCE(DEI) within the blue and orange groups, the variation of $\tilde{\eta}$ cannot be totally ascribed to the values of $\tau_\mathrm{p}$, $E_\mathrm{v}^{\mathrm{hl}}$, and $E_\mathrm{c}^{\mathrm{el}}$.
However, all the cases where ions severely reduce cell efficiency are included.

\subsection{Bias at MPP strongly influences mobile ion impact} \label{sec:bias}

\begin{figure}[tb]
    \centering
    \includegraphics[width=\linewidth]{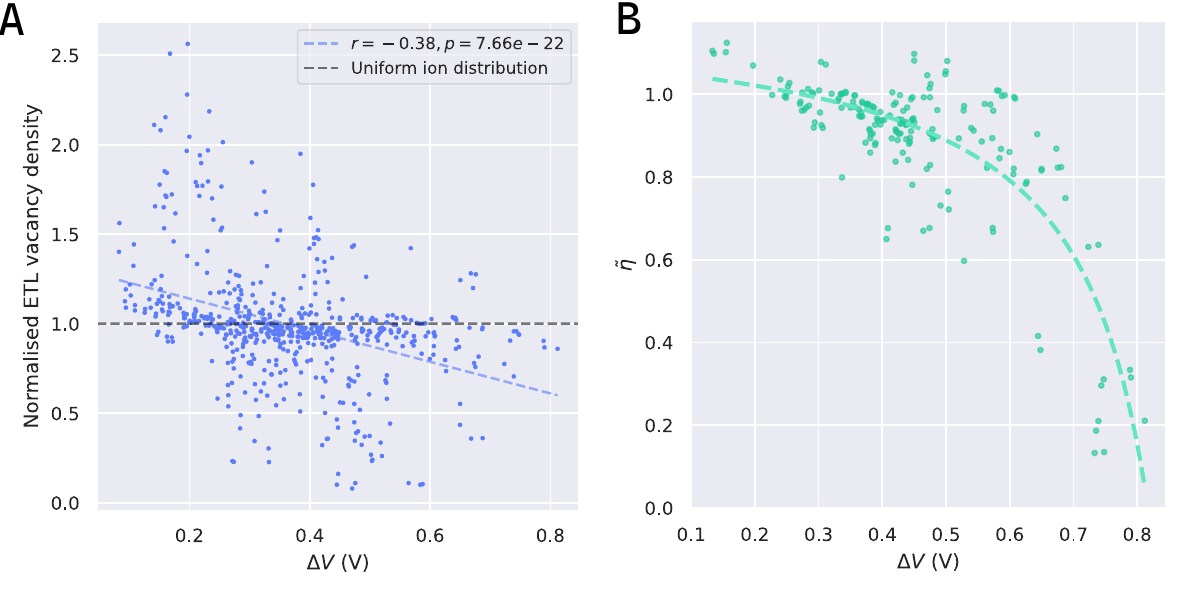}
    \caption{
    \textbf{Severe instances of negative mobile ion impact are due to a large built-in voltage relative to the maximum power point.}
    (A)
    The normalised vacancy density at the electron transport layer interface (ETL) plotted against the operational potential drop $\Delta V$.
    The blue dashed line shows a linear fit, with the Pearson correlation coefficient $r$ and corresponding $p$ value given in the legend.
    The black dashed line indicates the uniform ion distribution: the interfacial vacancy density at which there is no accumulation of vacancies in the device.
    (B) $\Delta V$ plotted against the ion-normalised PCE $\tilde{\eta}$ for devices with large built-in voltages and short hole lifetimes (devices coloured blue and orange in Figure~\ref{fig:correlations}).
    A rational function ($y=(ax+b)/(cx+d)$) fitted with an $R^2$ of 0.65 (dashed line).
    }
    \label{fig:vbivmpp}
\end{figure}

As we have seen above, certain low performing devices are very negatively impacted by the DEI case.
While we have identified several key conditions that are required for this to occur, namely a large built-in voltage combined with a short hole lifetime, we have not yet elucidated why devices with these conditions display a broad range of values of $\tilde{\eta}$, including some cells that are still improved when mobile ions are present.
The devices in this subset have a wide range of PCEs (orange and blue dots in Figure\,\ref{fig:correlations}B), spanning from $\approx$ 22 $\%$ to 5 $\%$.
We know from previous work that the field screening from charged ions is important, with the distribution of ions determining the impact it has \cite{Jacobs17,Garcia-Rodriguez22}.
A key point in this present study is that we are investigating the behaviour of each device at the maximum power point.
This point is device-specific and will result in different internal fields being present in each device, and therefore different ionic distributions.
This is because the applied voltage at the maximum power point relative to the built-in potential across the device will determine the ionic distribution \cite{Courtier18}.
To demonstrate this behaviour in our current dataset, we define a potential offset $\Delta V$ to be, 
\begin{equation} \label{eq:veff}
    \Delta V = V_{\mathrm{bi}} - V_{\mathrm{mpp}}.
\end{equation}
We calculate this quantity for the DEI devices and plot it against the vacancy density at the the ETL interface, normalised by the background density of mobile vacancies $N_I$ (Figure~\ref{fig:vbivmpp}A).
While there is a large degree of variation, it is clear from this plot that, on average, as $\Delta V$ increases, vacancy depletion occurs at the ETL.
The inverse (not presented) occurs at the HTL.
Therefore, the vacancy distribution during operation within a device depends to a large degree on the key quantity $\Delta V$.
This provides intuition for the impact of the band positions in Figure~\ref{fig:correlations}A, as these determine $V_{\mathrm{bi}}$, and thus the ionic distribution at $V_{\mathrm{mpp}}$.

Previous work has suggested that performance is degraded by halide vacancy accumulation at the HTL \cite{Jacobs17}, a situation that would be expected to occur at greater densities as $\Delta V$ is decreased.
Therefore, we hypothesise that if the maximum power point of the device is at a low voltage relative to the built-in voltage, the action of mobile ions at operational conditions will be to reduce performance.

To test this idea, we plot $\Delta V$ against $\tilde{\eta}$ for the devices identified in Figure~\ref{fig:correlations}B to have large built-in voltages and short hole lifetimes (Figure~\ref{fig:vbivmpp}B).
We find that there is a correlation between the number of volts that the maximum power point is below the built-in voltage and $\tilde{\eta}$.
For these devices, we propose a rational relationship between $\Delta V$ and $\tilde{\eta}$.
\begin{equation}
    \tilde{\eta} (\Delta V) = \frac{a\Delta V+b}{c\Delta V+d},
\end{equation}
where $a$, $b$, $c$, and $d$ are all unitless constants.
Fitting this relationship finds an $R^2$ of 0.65 and allows us to make several predictions about the relationships between $\Delta V$ and $\tilde{\eta}$ for the devices in this subset.
By rearranging our rational function for $\Delta V$ when $\tilde{\eta}=1$, such that,
\begin{equation}
    \Delta V |_{\tilde{\eta} = 1} = \frac{d-b}{a-c},
\end{equation}
we predict that, on average, any mobile ion concentration should have either a neutral or positive impact if we design a device to have $\Delta V <$ \textbf{0.27}\,V.
Finally, we can calculate the negative limit of this function via,
\begin{equation}
    \lim_{\Delta V\to-\infty} \tilde{\eta} (\Delta V) = \frac{a}{c},
\end{equation}
giving a value of $\tilde{\eta}=$\textbf{1.21}.
This sets an upper bound on the degree of improvement expected from increasing $\Delta V$, with the estimate being in line with the most ion-improved devices seen (Figure~\ref{fig:jv_statistics}A).

\section{Conclusions}

In this study we have investigated the impact of mobile ions on the steady state operational performance of PSCs across a range of parameter space. We observe that mobile ions can either improve or degrade steady state device performance depending on the material properties (Figure~\ref{fig:jv_statistics}).

Our primary conclusion, that mobile ions have minimal impact on high efficiency cells ($>20$\,\% PCE) is surprising, as the received wisdom of the field for a number of years has been that the presence mobile ions is a key driver of performance degradations. In our study we find that the two populations of 601 modelled devices, with and without mobile ions, are not significantly different enough to reject the null of the Brunner-Munzel test ($p = 0.10$). The high-performance devices modelled as part of this study are a mix of those with $N_I = 10^{17}$\,cm$^{-3}$ and $N_I = 10^{19}$\,cm$^{-3}$ (Figure~\ref{fig:ion_density_correlation}).
In many cases the `with mobile ion' PCE (DEI) is higher than the `no mobile ion' PCE (SUI), with SUI PCE $>20\%$ and $N_I = 10^{19}$\,cm$^{-3}$. This suggests highly efficiency devices can be made even in the presence of high concentrations of mobile ions.
Our factorial analysis allows us to identify the key parameters which are responsible for increases and decreases in performance in the studied parameter space. 

We identify the transport layer band positions and perovskite bulk hole lifetimes as having the greatest impact on $\tilde{\eta}$. Recombination has been discussed in the literature as a key driver of ion mediated impact, however, literature studies generally focus on recombination at the interface. In this study it is bulk recombination that has the biggest impact on efficiency. This may be a reflection of our recombination rate parameter choices.
While we have selected our parameter range based upon a through literature review, the sensitivity of $\tilde{\eta}$ to a parameter will be partly determined by the range selected. Thus, there may be regions of parameter space which identify a different recombination mechanism as key. However, we argue that the specific recombination process is not a key feature of the system, with the more general observation being that high rates of recombination from any source will result in increased impact of mobile ions \textit{when the built-in field is relied upon for extraction}.

The situation is more complex where mobile ions drive chemical reactions and irreversible materials degradation \cite{zhang_degradation_2022}, or where they create new recombination centres. Our results suggest that if mobile ions can be confined to the perovskite film and if only reversible chemical changes occur within the film, then the mere existence of mobile ions is not automatically detrimental to PSK devices, and there are regions of parameter space where ions have a beneficial effect on device performance. Our results suggest (\S~\ref{sec:vbi_recomb} \& \S~\ref{sec:bias}) it is for less efficient devices ($<15$\,\% PCE) that the presence of mobile ions is seriously deleterious.

It should be emphasised that our approach of simulating `ion free' SUI devices at steady state is not equivalent to experiments using fast JV sweeps to approximate a device without mobile ions.
For fast scan rate experiments, the choice of preconditioning voltage is key as the mobile ions will only be a non-participant if the preconditioning voltage results in a completely uniform ion distribution \cite{Cachafeiro25}.
As shown by Figure~\ref{fig:vbivmpp}A, this is not a simple linear relationship, though there are experimental methods that can provide good approximations of the voltage drop across the perovskite film \cite{Hill23, Hart24}. If the preconditioning voltage used is above $V_{\mathrm{bi}}$, then a fast JV sweep will report an artificially large PCE due to vacancy accumulation at the ETL driving increased extraction \cite{Cachafeiro25}.

As a final point, the mechanism by which mobile ions can severely affect steady state PCEs was examined. Here we identify a strong relationship between a large operating potential offset $\Delta V$ and reduced performance when mobile ions are introduced. One caveat to note is that in this work we use a simple approximation, that the flat-band potential is close to the built-in voltage as determined by our transport layer work functions. However, this is not completely accurate as non-symmetrical transport layer offsets and transport layer properties such as doping densities can also alter the flat-band potential.
In future work, our factorial approach can be applied to understanding how the complex parameter interactions of a multi-layer PSC influences the flat ion potential, and improve estimates of its value.

In conclusion, this study uses a diverse set of simulated PSC devices to investigate the effect of mobile ions on operational steady state, finding that within the material variation inspected here, there is no significant effect.
Using a factorial analysis we identify the key parameters determining which devices do see an impact from mobile ions, showing that there can be both negative and positive effects, though most of these are restricted to a change in PCE of $\pm$10\,\%
Finally, the position of $V_{\mathrm{mpp}}$ relative to $V_{\mathrm{bi}}$ is identified as a key indicator of whether the mobile ion field screening will negatively effect the device, providing an intuition of the device physics at play, and suggesting an important design guideline for engineering devices that are positively or neutrally impacted by mobile defects.

\subsection*{Conflict of Interest}
The authors declare that they have no conflict of interest.

\subsection*{CRediT} MVC carried out this work during his PhD, which was supervised by PJC (lead supervisor) and co-supervised by ABW and MJW. MVC conceived the study in discussion with the supervisory team; the simulations were carried out by MVC. The initial manuscript was written by MVC and modified by PJC with input from ABW, KOJ and MJW who also read, commented on and suggested changes to the draft. The authors would like to thank Dr. Ben Morgan (Uni of Bath) for feedback and suggestions relating to the modelling work. We also thank Dr. Will Clarke and Prof. Giles Richardson (Uni of Southampton) for always being ready to answer questions relating to IonMonger.

\bibliography{libraryABW.bib}%

\clearpage

\appendix
\section{Supplementary Information}
\renewcommand\thefigure{\thesection.\arabic{figure}}
\setcounter{figure}{0}

\subsection{Figures}

\begin{figure}[hbtp]
    \centering
    \includegraphics[width=0.44\linewidth]{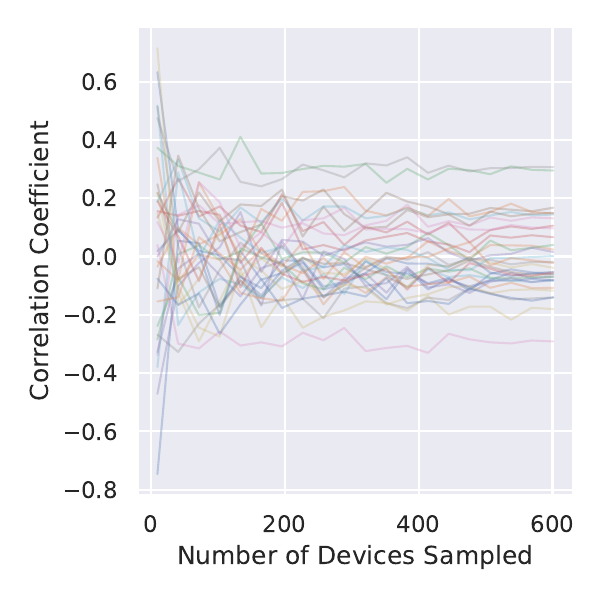}
    \caption{
    Line plot showing the convergence of Pearson's correlation coefficient between each model parameter (Table~\ref{tab:param_tab}) and the ion-normalised PCE $\tilde{\eta}$.
    }
    \label{fig:convergence}
\end{figure}

\begin{figure}[hbtp]
    \centering
    \includegraphics[width=0.44\linewidth]{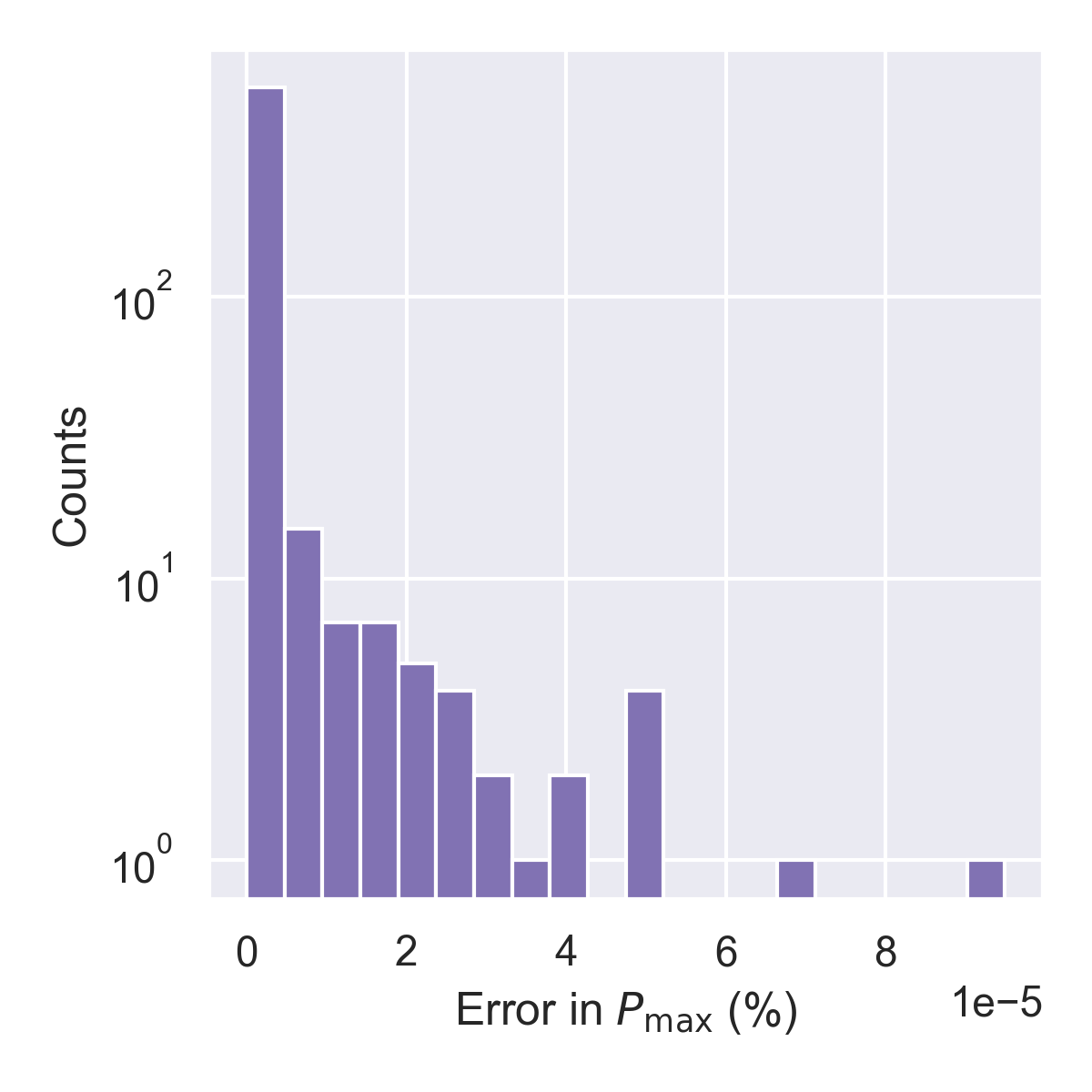}
    \caption{
    Histogram of the percentage error in the $P_{\mathrm{max}}$ found by ultra-slow 1$\times$10$^{-5}$ Vs$^{-1}$ compared with a simulation for each device where the VMPP is held static for 1$\times$10$^{-5}$s.
    }
    \label{fig:p_max_error}
\end{figure}

\begin{figure}[hbtp]
    \centering
    \includegraphics[width=0.44\textwidth]{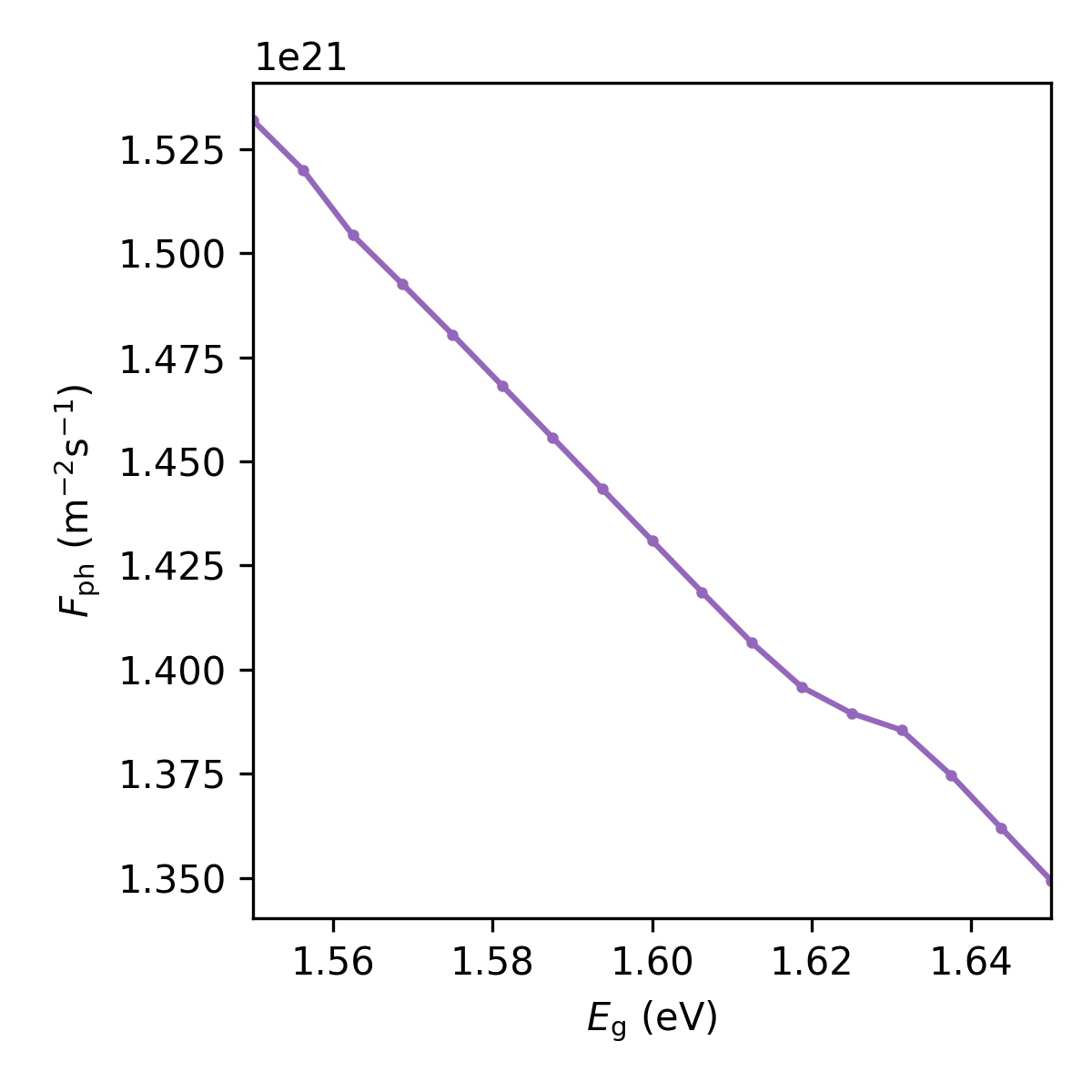}
    \caption{Plot showing how the simulation photon flux $F_{\mathrm{ph}}$ varies with the band gap $E_g$.}
    \label{fig:eg_fph}
\end{figure}

\begin{figure}[hbtp]
    \centering
    \includegraphics[width=0.44\linewidth]{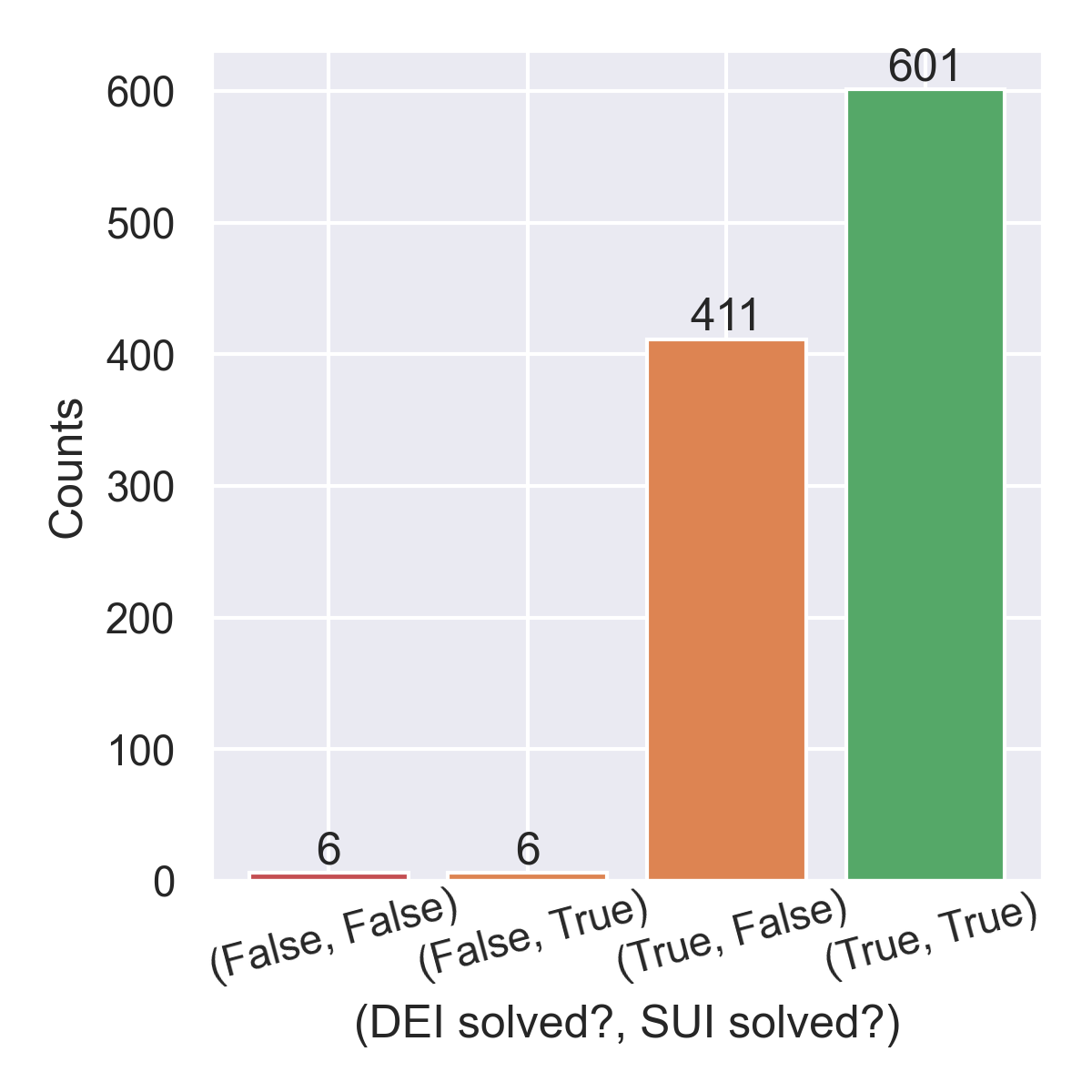}
    \caption{
    Bar plot showing the number of devices with numerically stable solutions in the DEI and SUI cases.
    }
    \label{fig:solved_bar_plot}
\end{figure}

\begin{figure}[hbtp]
    \centering
    \includegraphics[width=0.44\linewidth]{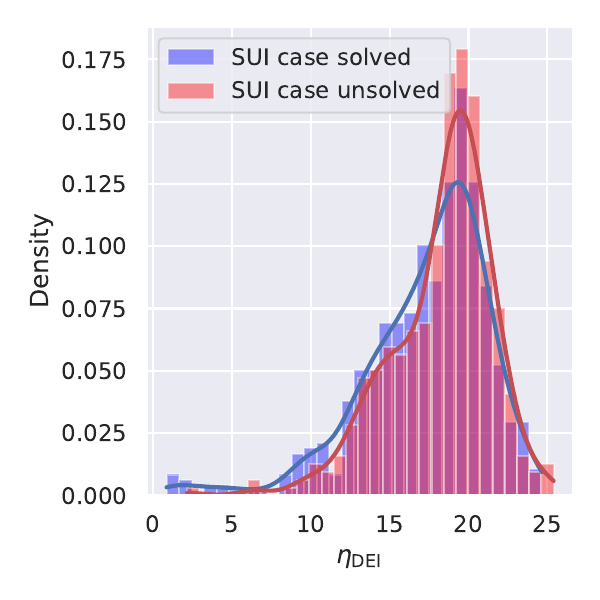}
    \caption{
    Kernel density estimation and histogram of the distribution of PCEs for the DEI case $\eta_{\mathrm{DEI}}$ from the devices included in the analysis, where both the DEI case and SUI case were solved numerically by IonMonger (blue), and also the excluded devices where the SUI case could not be solved (red).
    }
    \label{fig:kde_failed}
\end{figure}

\begin{figure}[hbtp]
   \centering
   \includegraphics[width=0.44\linewidth]{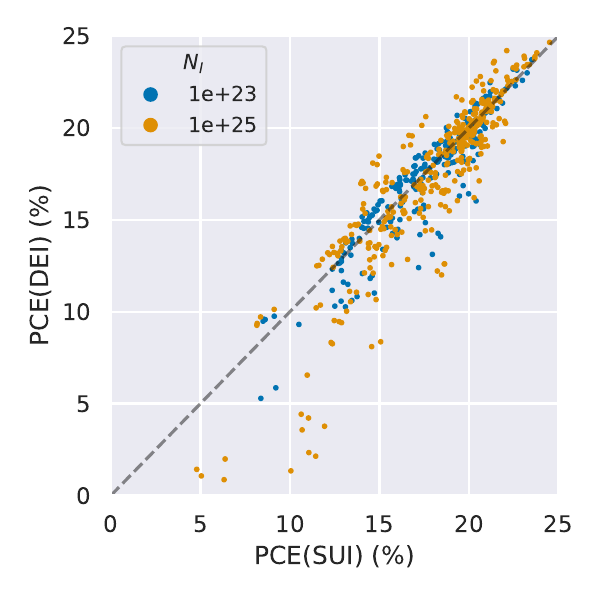}
   \caption{
   Scatter plot of power-conversion efficiencies (PCE) of parameter pairs of devices with mobile ions (DEI) and those without (SUI).
   Dots are coloured by the device equilibrium ion density $N_I$.
   }  \label{fig:ion_density_correlation}
\end{figure}

\clearpage

\subsection{Computing implementation}

A remote-access Linux-based server was utilised with 96 cores and 256\,GB of RAM.
Using this resource, a parallelised workflow was created using \texttt{multiprocessing} and the \texttt{MATLAB} Engine API for \texttt{Python}.
A bonus of this workflow, in addition to the computational speed-up, is the \texttt{timeout} argument of \texttt{multiprocessing}'s \texttt{AsyncResult.get()} method.
IonMonger solves many input parameter sets efficiently, but may struggle when given certain material properties and experimental conditions.
Having a timeout functionality allows for difficult-to-solve devices in the dataset to be skipped without blocking subprocesses from moving onto the next device in the job queue, accelerating the computation.


\end{document}